\documentclass[twocolumn,amstext,amsmath,amssymb,prl,superscriptaddress,floatfix,showpacs,nofootinbib]{revtex4}

\usepackage[dvips]{epsfig}
\usepackage{slashed}
\usepackage{mathrsfs}
\usepackage{graphicx}
\usepackage{xcolor}
\usepackage{hyperref}
\usepackage{dcolumn}
\usepackage{subfigure}
\usepackage{xspace}
\usepackage{bbm}

\usepackage{color}
\usepackage{ulem}

\newcommand{\mm}{\marginpar{\colorbox{green}{\textbf{BJ}}\\@Mario:}}
\newcommand{\bj}{\marginpar{\colorbox{green}{\textbf{MM}}\\@BJ:}}

\def\roughly#1{\mathrel{\raise.3ex\hbox{$#1$\kern-.75em%
\lower1ex\hbox{$\sim$}}}}

\def\g2k{\Gamma^{(2)}_k}

\def\ma0{m_{a_{0}}}
\def\mf0{m_{f_{0}}}

\definecolor{heidelbeer}{rgb}{0.5,0,0.5}

\graphicspath{
{./}
{../figures/}
}
\begin{document}
\title{Spin Effect induced Momentum Spiral and Asymmetry Degree in Pair Production}
\author{Li-Na Hu}
\affiliation{Key Laboratory of Beam Technology of the Ministry of Education, and College of Nuclear Science and Technology, Beijing Normal University, Beijing 100875, China}
\author{Hong-Hao Fan}
\affiliation{Key Laboratory of Beam Technology of the Ministry of Education, and College of Nuclear Science and Technology, Beijing Normal University, Beijing 100875, China}
\author{Orkash Amat}
\affiliation{Key Laboratory of Beam Technology of the Ministry of Education, and College of Nuclear Science and Technology, Beijing Normal University, Beijing 100875, China}
\author{Suo Tang}
\affiliation{College of Physics and Optoelectronic Engineering, Ocean University of China, Qingdao, Shandong 266100, China}
\author{Bai-Song Xie \footnote{bsxie@bnu.edu.cn}}
\affiliation{Key Laboratory of Beam Technology of the Ministry of Education, and College of Nuclear Science and Technology, Beijing Normal University, Beijing 100875, China}
\affiliation{Institute of Radiation Technology, Beijing Academy of Science and Technology, Beijing 100875, China}
\date{\today}

\begin{abstract}
Spin effect on the pair production under circularly polarized fields are investigated. Significantly different from what momentum spirals caused by two counter-rotating fields with a time delay, we find for the first time that the spirals can also be induced due to the particles spin effect even if in a single field. We further examine the bichromatic combinational fields, the inhomogeneous spiral structures can be observed in the momentum spectrum, in particular, the spiral not only does exist in two cases of spin but also is about two orders of magnitude amplifier than that in the single field. Meanwhile, the spin asymmetry degree on the momentum distributions is investigated and found that there exist the effect of spin flip with increasing time delay between two fields. The spin asymmetry degree on the number density can reach to $98\%$ in a certain of condition. These results indicate that the signatures of created particles, especially the spiral structures are strongly associated with the information of laser field as well as the created particle spin, which can deepen the understanding of vacuum pair production.
\end{abstract}

\maketitle

\textit{Introduction.}--In the presence of strong background field, the quantum electrodynamic (QED) vacuum is unstable and decays into electron-positron pairs \cite{Sauter:1931zz,Heisenberg:1935qt,DiPiazza:2011tq,Fedotov:2022,Xie:2017,
Dunne:2008kc,Schwinger:1951nm,Burke:1997ew}. A great deal of research has focused on the momentum distribution and number density of created pairs for different external fields. The momentum distribution contains rich features such as effective mass signatures \cite{Kohlfurst:2013ura}, the self-bunching effect \cite{Hebenstreit:2011wk}, the ponderomotive force effect \cite{Kohlfurst:2017hbd}, the multiple-slit interference effect \cite{Li:2014En}, node structures \cite{Li:2015Non}, as well as recently discovered spiral structures \cite{Li:2017qwd,Hu:2023pmz,Li:2019hzi}.

In fact, spiral structures have been widely investigated in many areas in past years, for example, in atomic and molecular ionization \cite{Ngoko:2015Ele,Pengel:2017Elec,Majczak:2022,Macek2009,K2016,L2020}, nonlinear optics \cite{Harris1994}, type-II superconductors \cite{Blatter1994}, plasmas physics \cite{Uby1995,Shukla2009}, atomic condensates \cite{Madison2000}, and so on. Interestingly, recent investigations indicate that the momentum spirals are discovered in electron-positron pair production \cite{Li:2017qwd,Hu:2023pmz,Li:2019hzi}, which are generated by two counter-rotating fields with a time delay. It should be noted that the spirals with even and odd arms are attributed to the fields by the same color \cite{Li:2017qwd,Hu:2023pmz} and by the different colors \cite{Li:2019hzi}, respectively.

On the other hand, it is known that the spin effect of created particles plays a critical role in vacuum pair production \cite{Strobel:2014tha,Blinne:2015zpa,Li:2019,Kohlf2019S}. For example, Strobel {\it et al.} proved that the spin distribution of produced pairs is generally not $1:1$ in the Schwinger pair production by the rotating electric fields depending on time \cite{Strobel:2014tha}. Blinne {\it et al.} pointed out that the pair production in different spin states is unequal for constant rotating field \cite{Blinne:2015zpa}. Interestingly, we found for circularly polarization (CP) field that the number density of created bosons is the geometric mean, i.e., the square root of spin-up number density times that of spin-down \cite{Li:2019}, meanwhile, for arbitrarily polarized multicycle fields,  that the degree of spin polarization roughly increases with $\gamma\ll1$, where $\gamma=m\omega/eE_{0}$ is Keldysh adiabatic parameter, $m$ and $-e$ denote electron mass and charge, respectively. Recently, Kohlf\"{u}rst showed the power of the photon absorption model by demonstrating how to identify the imprint the different spin states leave on the electron-positron angular distribution, and obtained the spin-dependent particle production amplitudes \cite{Kohlf2019S}.

To our knowledge, however, whether the spin of created particles can create momentum spiral has not yet been reported. Therefore, as an important concern, we shall investigate the problem in this Letter. The focus is on considering the spin induced momentum spiral by the single field, and the spiral structure as well as the spin asymmetry degree under the bichromatic combination fields. For the single field, it is found that there is no momentum spiral when the spin of created particles is parallel to the direction of laser field ($s=+1$). Surprisingly, when the spin is antiparallel to the field direction ($s=-1$), there exists an obvious spiral structure in the momentum spectrum. And the spiral in the subcycle field is more remarkable than that in the multicycle field. On the other hand, for the bichromatic combination fields, the spiral does exist no matter what is for the $s=+1$ or/and $s=-1$. Not only the unobserved spiral in a single field is seen in the two fields, but also the observed spiral in a single field is amplified about two orders of magnitude larger in the bichromatic fields. Meanwhile, it is found that the larger the time delay between the two fields, the more pronounced the spiral structure. Finally, by adjusting appropriate field parameters, we can also obtain high spin asymmetry degree of created particles.

Our considered electric field form is associated with either of one single field or their combination by two
fields with a little different strength and frequency between them. For single field, it is $\mathbf{E}(t)=\mathbf{E}_{1}(t)=\frac{E_1}{\cosh(\frac{t}{\tau_1})}[\cos(\omega_1 t)\mathbf{e}_x+\delta_1\sin(\omega_1 t)\mathbf{e}_y]$ or $\mathbf{E}(t)=\mathbf{E}_{2}(t)=\frac{E_2}{\cosh(\frac{t-T_d}{\tau_2})}[\cos(\omega_2(t-T_d))\mathbf{e}_x+ \delta_2\sin(\omega_2(t-T_d))\mathbf{e}_y]$. And for the combined field case, it is $\mathbf{E}(t)=\mathbf{E}_{1}(t)+\mathbf{E}_{2}(t)$, where $E_{1,2}=E_{01,02}/\sqrt{1+\delta_{1,2}^2}$ are the field strengths ($|\delta_{1,2}|=1$ are the CP degree), $\omega_{1,2}$ are the field frequencies, $\tau_{1,2}=N2\pi/\omega_{1,2}$ denote the pulse duration ($N$ determines the number of cycles in the individual pulse), $T_d$ is the time delay between the two pulses (it is in unit of $T=\tau_1+\tau_2$). In our study, the typical field parameters are set as $E_{01}=0.1\sqrt{2}E_{cr}$, $E_{02}=0.07\sqrt{2}E_{cr}$, $\delta_{1}=-1$, $\delta_{2}=1$, $\omega_1=0.44$, $\omega_2=0.55$. Note that $\hslash=c=1$ are used and all quantities are presented in terms of the electron mass $m$.

We employ the equal-time Dirac-Heisenberg-Wigner (DHW) formalism \cite{Strobel:2014tha,Blinne:2015zpa,Kohlfurst:2015zxi,Hebenstreit:2011pm,Bialynicki-Birula:1991jwl,Hebenstreit:2010vz}, a relativistic phase-space quantum kinetic approach, which has been widely applied to study pair production in arbitrary electromagnetic fields \cite{Ababekri:2019dkl,Kohlf2020Effect,Blinne:2013via,Li:2021vjf,Hebenstreit:2011wk,Olugh:2018seh,
Li:2015cea,Kohlfurst:2017git,Olugh:2019nej}. For the spatially homogeneous and time-dependent electric field presented in our study, the complete DHW equations of motion can be reduced to the following form \cite{Blinne:2015zpa,Li:2017qwd,Olugh:2019nej}
\begin{equation}
\begin{split}
\dot{f}&=\frac{e\mathbf{E}\cdot \mathbf{v}}{2\Omega},\\
\dot{\mathbf{v}}&=\frac{2}{\Omega^{3}}[(e\mathbf{E}\cdot \mathbf{p})\mathbf{p}-e\mathbf{E}\Omega^{2}](f-1)-\frac{(e\mathbf{E}\cdot \mathbf{v})\mathbf{p}}{\Omega^{2}}\\
&-2\mathbf{p}\times \mathbbm{a}-2m\mathbbm{t},\\
\dot{\mathbbm{a}}&=-2\mathbf{p}\times \mathbf{v},\\
\dot{\mathbbm{t}}&=\frac{2}{m}[m^{2}\mathbf{v}+(\mathbf{p}\cdot \mathbf{v})\mathbf{p}],
\end{split}\label{eq3}
\end{equation}
with initial conditions $f(\mathbf{q},-\infty)=0$, $\mathbf{v}(\mathbf{q},-\infty)=\mathbbm{a}(\mathbf{q},-\infty)=\mathbbm{t}(\mathbf{q},-\infty)=0$, where $f$ is the single-particle momentum distribution function, $\mathbf{v}$ is associated with the current density $\mathbbm{v}$, $\mathbbm{a}$ denotes spin density, and the projection of the $\mathbbm{t}$ in the direction of momentum determines the rate of change of the mass density \cite{Kohlfurst:2015zxi,Hebenstreit:2011pm,Kohlf2019S}. $\Omega=\sqrt{m^{2}+(\mathbf{q}-e\mathbf{A}(t))^{2}}$ represents the total energy of particles, $\mathbf{q}$ is the canonical momentum related to the kinetic momentum $\mathbf{p}(t)=\mathbf{q}-e\mathbf{A}(t)$, $\mathbf{A}(t)$ is the vector potential of the electric field $\mathbf{E}(t)$.
By integrating the $f(\mathbf{q},t)$ over full momenta at $t\rightarrow+\infty$, we can obtain the number density of created particles:
\begin{equation}\label{14}
  n = \lim_{t\to +\infty}\int\frac{d^{3}q}{(2\pi)^ 3}f(\mathbf{q},t) \, .
\end{equation}

Note that the above $f$ and $n$ are the forms without considering the spin of created particles. When considering the spin, the single-particle momentum distribution function can be changed as \cite{Blinne:2015zpa,Li:2019}
\begin{equation}\label{15}
 f_s= \frac{1}{2} \left( f+s \, \delta f_\mathrm{sc} \right),
\end{equation}
where $s=\pm1$, $\delta f_\mathrm{sc}= \frac{q_z}{\epsilon_\perp}\delta f_\mathrm{c}+\frac{m}{\epsilon_\perp}\delta f_{\mu_z}\,$ is associated with the chiral asymmetry $\delta f_\mathrm{c}= \frac{1}{2\Omega}\mathbf{p}\cdot\mathbbm{a}$ and the magnetic moment asymmetry $\delta f_\mathrm{\mu_z} = \frac{1}{2\Omega}\left( m a_z+(\mathbf{p}\times\mathbbm{t}\,)_z \right)$,
$\epsilon_\perp=\sqrt{m^2+{q_z}^2}$. Also the corresponding number density of considering the spin could be modified as $n_s= \lim_{t\to +\infty}\int\frac{d^{3}q}{(2\pi)^ 3}f_s(\mathbf{q},t) \,$.

\begin{figure}[htbp]
\begin{center}
\includegraphics[width=0.48\textwidth]{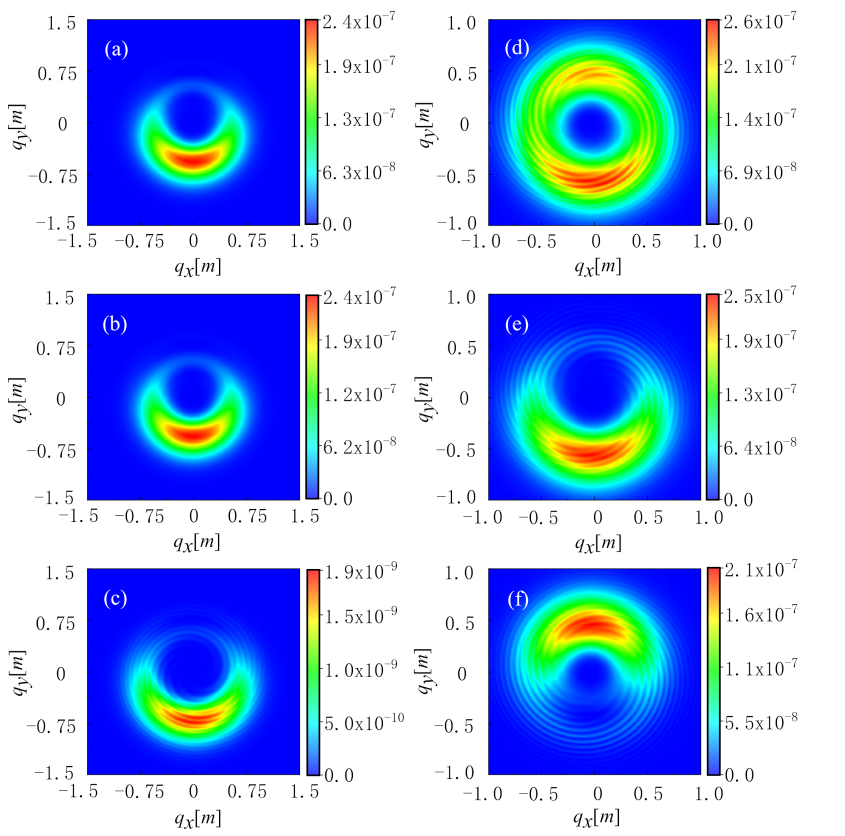}
\end{center}
\vspace{-0.01cm}
\setlength{\abovecaptionskip}{-0.05cm}
\setlength{\belowcaptionskip}{-0.3cm}
\caption{(color online). Momentum spectra of created particles in the ($q_{x}$, $q_{y}$) plane (where $q_{z}=0$) under the single field $\mathbf{E}_{1}(t)$ (for the first column) and the bichromatic combination fields for $T_d=8T$ (for the second column) with $N=0.5$. From up to down, each row corresponds to $f(q_x, q_y)$, $f_{+}(q_x, q_y)$ and $f_{-}(q_x, q_y)$, respectively. Other parameters are $E_{01}=0.1\sqrt{2}E_{\mathrm{cr}}$, $\delta_{1}=-1$, $\omega_{1}=0.44$, $E_{02}=0.07\sqrt{2}E_{\mathrm{cr}}$, $\delta_{2}=1$, $\omega_{2}=0.55$.}
\label{1}
\end{figure}

\textit{Momentum distribution.}--The momentum spectra of created particles in the single field $\mathbf{E}_{1}(t)$ and the bichromatic combination fields with $N=0.5$ are shown in Fig.~\ref{1}. It is found that under the $\mathbf{E}_{1}(t)$, for $f(q_x, q_y)$, there is no obvious spiral structure but the distribution presents a good symmetry in the $q_x$ direction, see Fig.~\ref{1}(a). The symmetry is related to the fact that the electric field in the $x$ direction is an even function \cite{Li:2015Non,Hu:2023pmz}. Meanwhile, one can see that the $f(q_x, q_y)$ is mainly distributed in the lower half plane, where it dominates the contribution to the pair production, which is consistent with the results in terms of the turning points by $\Omega(\mathbf{q}, t)=0$ \cite{Dumlu1011,Strobel2015}. For $f_{\pm}(q_x, q_y)$, however, the momentum distributions present a clear difference. We find that the $f_{+}(q_x, q_y)$ shown in Fig.~\ref{1}(b) is dominant. But surprisingly, there exists a remarkable spiral structure in the $f_{-}(q_x, q_y)$, see Fig.~\ref{1}(c), although it is about two orders of magnitude weaker than the $f_{+}(q_x, q_y)$. Actually, we do the results of $\mathbf{E}_{2}(t)$ (patterns are not shown), which are similar to the results of $\mathbf{E_1}(t)$, except that the momentum distribution shifts from the lower half plane to the upper half plane, and the spiral structures appears in the $f_{+}(q_x, q_y)$. These are not surprising since that the spin directions of the fields $\mathbf{E}_{1}(t)$ and $\mathbf{E}_{2}(t)$ are opposite with each other.

Note that when we consider the spin of created particles, there are two ways for spin alignment in a pair of spin-1/2 particles: parallel and antiparallel \cite{Kohlf2019S}. For the parallel, the total spin is $S=1/2+1/2=1$, while for the antiparallel, the total spin is $S=0$. Obviously, it is found that the distribution of created pairs for $S=1$ is larger than that for $S=0$, which results in agreement with that in Ref.\cite{Kohlf2019S}.

Now let us discuss briefly the possible relation for the distribution region, for example it is mainly focused in the lower half plane when field $\mathbf{E}_{1}(t)$ is applied, to the conservation of total angular momentum in pair production process. Note that the total angular momentum is composed of orbital angular momentum (OAM) and spin angular momentum, i.e., $J=L+S$. It is known that the electron-positron pairs are mainly produced in about a field-period near the $t=0$ since that the field strength is larger than those in other time regime. For one pair of electron-positron, the OAM produced in this period has a positive trend fortunately when the momentum is located in lower half plane. Therefore, in the probability sense, we can think the OAM is positive, which is indeed required for guaranteeing the conservation of total angular momentum in pair production precess. For example, if one-pair electron-positron is produced via absorbing at least $5$ photons (since the original central frequency is $\omega_1=0.44$), then the OAM of electron/positron should be $L=2$ when $S=1$ so that the total angular momentum is $J=5$. Certainly exact descriptions and examinations about the total angular momentum conservation problem need to adopt the more powerful theoretical tool such as the S-matrix treatment.

It should be emphasized that the momentum spirals in a single field in our study mentioned above are induced by the spin effect of created particles, which is different from the previous one \cite{Li:2017qwd,Hu:2023pmz,Li:2019hzi} that is caused by the opposite photons' spin in two counter-rotating fields with a time delay.

Under the bichromatic combination fields with $T_d=8T$, it is found that there are the remarkable spiral pattern in the momentum spectra. For $f(q_x, q_y)$, the spiral structure is constituted of nine arms and its distribution in the whole plane is nonuniform with a stronger in lower half plane, see Fig.~\ref{1}(d). This phenomenon can be understood in connection with the results of a single field. It is known that the momentum spectra $f(q_x, q_y)$ of $\mathbf{E}_{1}(t)$ and $\mathbf{E}_{2}(t)$ are mainly distributed in the lower half plane and the upper half plane, respectively. And due to the field strength of $\mathbf{E}_{1}(t)$ is larger than that of $\mathbf{E}_{2}(t)$, so the $f(q_x, q_y)$ of $\mathbf{E}_{1}(t)$ is stronger than ones of $\mathbf{E}_{2}(t)$. When time delay is large, the two fields $\mathbf{E}_{1}(t)$ and $\mathbf{E}_{2}(t)$ are almost independent, which leads to the result under the combined fields being almost a superposition of the ones under the two individual fields. Therefore, for $T_d=8T$, the $f(q_x, q_y)$ of the lower half plane is stronger than that of the upper half plane.

For $f_{\pm}(q_x, q_y)$, one can see that the spiral of $f_{+}(q_x, q_y)$ is mainly distributed in the lower half plane, as shown in Fig.~\ref{1}(e), while the result of $f_{-}(q_x, q_y)$ is opposite, see Fig.~\ref{1}(f). The results are also associated with the momentum distributions in a single field. Specifically, let's first understand the result of $f_{+}(q_x, q_y)$ in Fig.~\ref{1}(e). In the combined fields with larger time delay, we can think that the $f_{+}(q_x, q_y)$ is consisted of the $f_{+}(q_x, q_y)$ of $\mathbf{E}_{1}(t)$ and $\mathbf{E}_{2}(t)$. It is known that the $f_{+}(q_x, q_y)$ of $\mathbf{E}_{1}(t)$ and $\mathbf{E}_{2}(t)$ are predominantly distributed in the lower half-plane and the upper half-plane, respectively, and there exists an obvious spiral structure in the $f_{+}(q_x, q_y)$ of $\mathbf{E}_{2}(t)$. Meanwhile, the $f_{+}(q_x, q_y)$ of $\mathbf{E}_{1}(t)$ is about two orders of magnitude larger than that of $\mathbf{E}_{2}(t)$. When the $\mathbf{E}_{1}(t)$ and $\mathbf{E}_{2}(t)$ are combined, there should be $f_{+}(q_x, q_y)$ in the whole plane. But since the $f_{+}(q_x, q_y)$ of $\mathbf{E}_{1}(t)$ mainly contributes to the position of distribution, while the $f_{+}(q_x, q_y)$ of $\mathbf{E}_{2}(t)$ mainly contributes to the spiral structure. Therefore, these results ultimately lead to the $f_{+}(q_x, q_y)$ in the combined fields mainly distributed in the lower half plane and accompanied by a remarkable spiral structure, where the spiral is about two orders of magnitude amplifier than that in the single field. For the results of $f_{-}(q_x, q_y)$ in the combined fields, a similar understanding can be performed.

\begin{figure}[htbp]
\begin{center}
\includegraphics[width=0.48\textwidth]{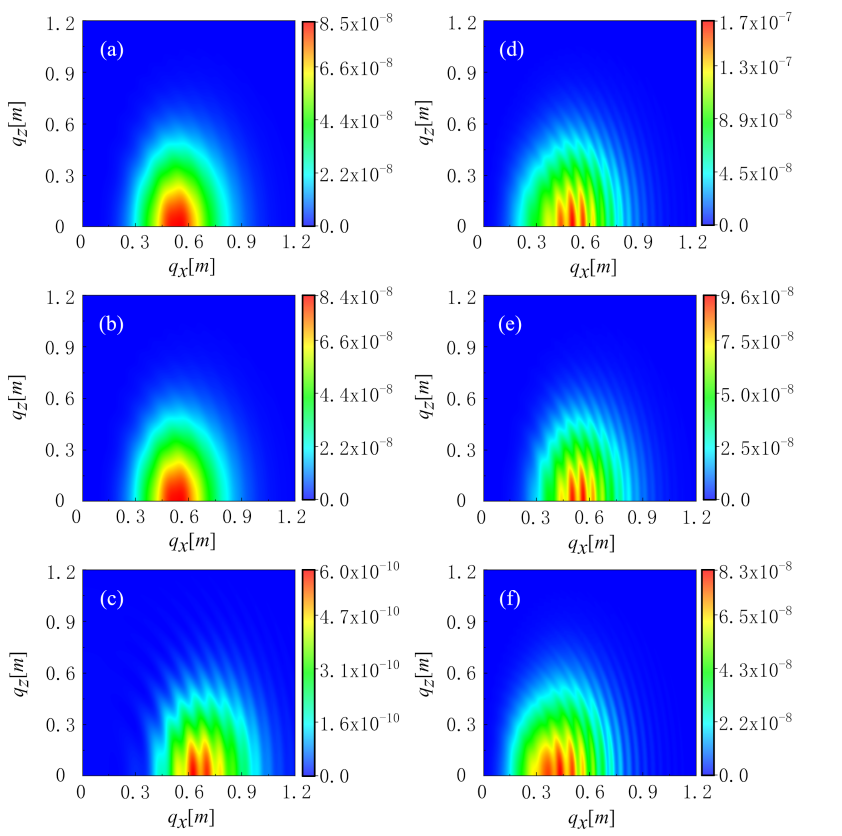}
\end{center}
\vspace{-0.01cm}
\setlength{\abovecaptionskip}{-0.05cm}
\setlength{\belowcaptionskip}{-0.3cm}
\caption{(color online). The same as in Fig.~\ref{1} except that the distributions are in the ($q_{x}$, $q_{z}$) plane (where $q_{y}=0$).}
\label{2}
\end{figure}

\begin{figure*}[htbp]
\begin{center}
\includegraphics[width=0.9\textwidth]{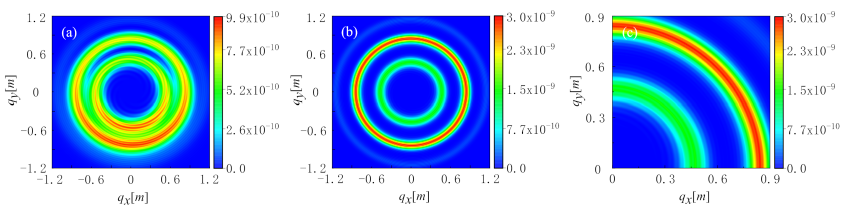}
\end{center}
\vspace{-0.01cm}
\setlength{\abovecaptionskip}{-0.1cm}
\setlength{\belowcaptionskip}{-0.5cm}
\caption{(color online). Momentum spectra $f_{-}(q_x, q_y)$ of created particles in the ($q_{x}$, $q_{y}$) plane (where $q_{z}=0$) for the single field $\mathbf{E}_{1}(t)$ with different $N$. (a) and (b) correspond to $N=1$ and $N=2$, respectively. (c) is the amplified result of (b).}
\vspace{0.2cm}
\label{3}
\end{figure*}

Now let us to see the momentum distributions in the ($q_{x}$, $q_{z}$) plane (where $q_{y}=0$) under the single or/and the bichromatic combined fields with $N=0.5$, the results are shown in Fig.~\ref{2}. Due to the distributions are symmetric about both of $q_{x}=0$ and $q_{z}=0$, we only show a quarter of the results here. Compared to the results in the previous work \cite{Blinne:2015zpa} for the single field, our $f_{-}(q_x, q_z)$ is similar to that $f_{+}$ of Fig.5 in \cite{Blinne:2015zpa} since the field is opposite to ours so that our $f_{-}$ corresponds to $f_{+}$ of them. However, our $f(q_x, q_z)$ and $f_{+}(q_x, q_z)$ are different from them. We think the differences may be attributed to that the field of \cite{Blinne:2015zpa} is for the multicycle case, and ours is for the subcycle case. In \cite{Blinne:2015zpa} they emphasized that the pattern is a typical interference effect, however, in our case here we have observed a delicate relation of interference to the weak spiral. For the combination fields, see the right column of Fig.~\ref{2},  similar to the results in the ($q_{x}$, $q_{y}$) plane of Fig.~\ref{1}, all momentum distributions have exhibited the interference or/and weak spiral including the magnitude amplifier effect (Fig.~\ref{2}f) compared to that in the single field (Fig.~\ref{2}c).

As $N$ increases from $N=0.5$ to $N=1$ and $N=2$, we take the results of $f_{-}(q_{x}, q_{y})$ in the $\mathbf{E}_{1}(t)$ as an example, see Fig.~\ref{3}, and find that the distribution range of the spiral becomes larger and the spiral arms become thinner as well as longer, while the symmetry of spiral becomes better in the both of $q_{x}$ and $q_{y}$ directions (To see more clearly, we amplify the result of $N=2$ and show a quarter of the ones.) These phenomena are associated with the fact that the pair production tends more and more to be the multiphoton absorption process with increasing $N$.

\begin{figure*}[htbp]
\begin{center}
\includegraphics[width=0.9\textwidth]{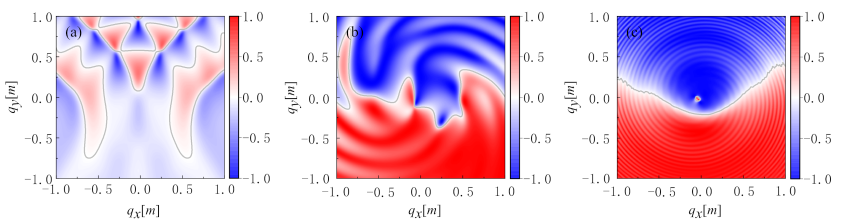}
\end{center}
\vspace{0.1cm}
\setlength{\abovecaptionskip}{-0.1cm}
\setlength{\belowcaptionskip}{-0.5cm}
\caption{(color online). Spin asymmetry degree on the momentum spectra in the ($q_{x}$, $q_{y}$) plane (where $q_{z}=0$) under the combination fields for different time delays with $N=0.5$. (a), (b) and (c) correspond to $T_d=0$, $T_d=T$ and $T_d=8T$, respectively. The gray line denotes $\kappa_{f}=0$.}
\label{4}.
\end{figure*}

\textit{Spin asymmetry degree.}--The spin asymmetry degree on the momentum distributions $f(q_{x}, q_{y})$ in the combined fields with different time delays are shown in Fig.~\ref{4}, in which there are the following three cases in general. (i) For $T_d=0$, the $f_{+}(q_x, q_y)$ and $f_{-}(q_x, q_y)$ are uniformly inter-nested in the whole plane, and the fraction of two distributions is almost $1:1$. (ii) For the small time delay $T_d=T$, there is a preliminary separation between the $f_{+}(q_x, q_y)$ and $f_{-}(q_x, q_y)$, but the $f_{-}(q_x, q_y)$ is dominant. (iii) For the large time delay $T_d=8T$, the $f_{+}(q_x, q_y)$ and $f_{-}(q_x, q_y)$ are completely separated, in which the upper half plane and the lower half plane are dominated by the $f_{-}(q_x, q_y)$ and $f_{+}(q_x, q_y)$, respectively.

For convenience, we define the spin asymmetry degree as
\begin{eqnarray}\label{asymmetry degree}
\kappa_{f}=\frac{f_{+}(q_x, q_y)-f_{-}(q_x, q_y)}{f_{+}(q_x, q_y)+f_{-}(q_x, q_y)}.
\end{eqnarray}
First, for $T_d=0$, one can see that the distributions of $\kappa_{f}$ has a good symmetry in the $q_x$ direction, see Fig.~\ref{4}(a), which is related to the symmetry of corresponding momentum distributions. Moreover, it is found that the $f_{+}(q_x, q_y)$ and $f_{-}(q_x, q_y)$ are uniformly inter-nested in Fig.~\ref{4}(a). This is because the the fraction of $f_{+}(q_x, q_y)$ and $f_{-}(q_x, q_y)$ is almost $1:1$. Second, as the time delay increases to $T_d=T$, the symmetry of $\kappa_{f}$ is destroyed, at the same time, a wide and nonuniform spiral structure can be observed, see Fig.~\ref{4}(b). In addition, there is preliminary separation between the $f_{+}(q_x, q_y)$ and $f_{-}(q_x, q_y)$, where the lower half plane is mainly dominated by the $f_{+}(q_x, q_y)$, while the results in the upper half plane are opposite. But in the whole plan, the $f_{-}(q_x, q_y)$ is dominant.

Third, as $T_d$ further increases to $T_d=8T$, the distribution of $\kappa_{f}$ is approximately symmetric in the $q_x$ direction, and we can observe the thin and uniform spiral fringes, see Fig.~\ref{4}(c). Moreover, it is found that there exists a clear dividing line in this distribution, where the lower half plane and the upper half plane are primarily dominated by the $f_{+}(q_x, q_y)$ and $f_{-}(q_x, q_y)$, respectively, which are associated with the ones of Figs.~\ref{1}(e) and (f). This clear dividing phenomenon may be attributed to the fact that when the $\mathbf{E}_{1}(t)$ and $\mathbf{E}_{2}(t)$ are combined, as we described above, the result with a large time delay is almost contributed by the superposition of the results in two individual fields.

\begin{figure}[htbp]
\begin{center}
\includegraphics[width=0.45\textwidth]{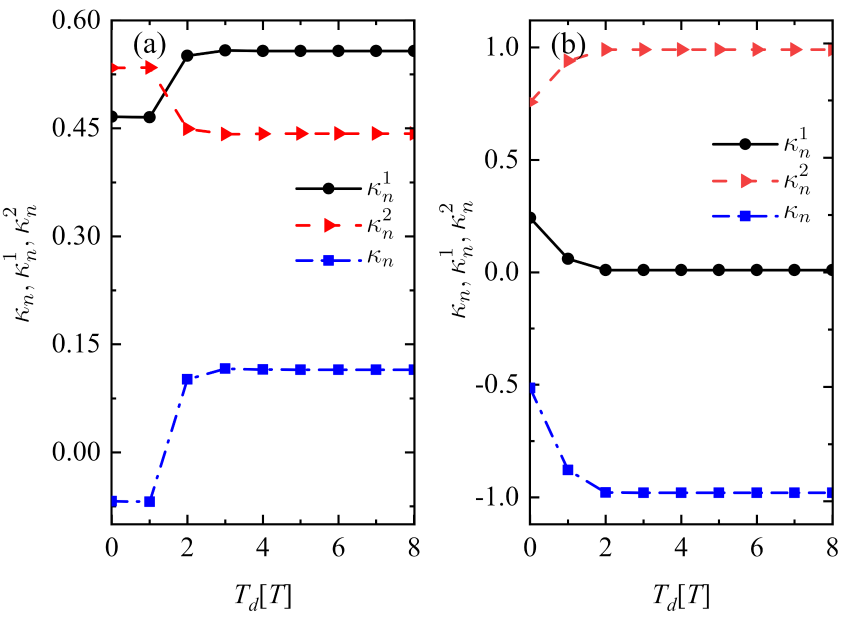}
\end{center}
\vspace{-0.01cm}
\setlength{\abovecaptionskip}{-0.2cm}
\setlength{\belowcaptionskip}{-0.8cm}
\caption{(color online). Spin asymmetry degree on the number density of created particles dependence on time delays by the combinational fields. (a) for $N=0.5$. (b) for $N=2$, $E_{02}=0.05\sqrt{2}E_{\mathrm{cr}}$ and $\omega_{2}=1.2$. Other parameters are the same as in Fig.~\ref{1}.}
\label{5}.
\end{figure}

The spin asymmetry degree on the number density of created particles in the combined fields with different time delays are shown in Fig.~\ref{5}, where we define the spin asymmetry degree as $\kappa_{n}=\kappa^{1}_{n}-\kappa^{2}_{n}$ with $\kappa^{1}_{n}=n_{+}/(n_{+}+n_{-})$ and $\kappa^{2}_{n}=n_{-}/(n_{+}+n_{-})$. From Fig.~\ref{5}(a), we find that for $T_d=0$, $\kappa^{1}_{n}$ is smaller than $\kappa^{2}_{n}$, i.e., the $n_{-}$ is dominant. As $T_d$ increases, a crossing point between $\kappa^{1}_{n}$ and $\kappa^{2}_{n}$ occurs, which is located in the range of $T_d=1\sim2T$. It indicates that there exists a spin-flip effect in the range. When $T_d\geq2T$, $\kappa^{1}_{n}$ is larger than $\kappa^{2}_{n}$, i.e., $n_{+}$ is dominant. Based on the above results, we can expect to control the momentum spectrum via changing the time delay, for example, if we want to obtain a momentum spectrum dominated by $f_{-}(q_x, q_y)$, we can set the time delay $T_d$ to be smaller. While if we want to get the spectrum dominated by $f_{+}(q_x, q_y)$, the $T_d$ needs to be extended. Moreover, one can see that with the increase of $T_d$, the $\kappa_{n}$ first increases and then remains unchanged as $12\%$. In order to obtain a higher spin asymmetry degree, we have performed some other simulations, for example the result shown in Fig.~\ref{5}(b). It is found that the $\kappa_{n}$ monotonically decreases with increasing $T_d$, specifically, it decreases from $-50\%$ at $T_d=0$ and eventually reaches to $-98\%$ at $T_d=8T$.

\textit{Discussion.}-- Different from the previous studies \cite{Li:2017qwd,Hu:2023pmz}, we found that the spin can induce the momentum spirals in a single field. The bichromatic combinational fields lead to an inhomogeneous spiral structures of the momentum, especially, the spiral pattern does exist in both of $f_{+}$ and $f_{-}$, and it is about two orders of magnitude larger than the single field. It seems that one field provides to create spiral "seed" for a certain spin type, while the other field plays a role as the "amplifier" to this spiral. For the opposite spin case, the role of "seed" and "amplifier" played by the fields is exchangeable.

It is known that the purely time-dependent electric field actually consists of two counter-propagating waves \cite{Kohlfurst:2022edl} and they are focused on the narrow regime of original electron Compton scales, i.e., a dipole approximation is met. For our single field, it is equivalent to a combination of a left-handed wave and a right-handed wave, and the two counter-propagating waves are attributing for most of the interaction time at the focusing spatial scale. So this is very similar to the case of combined fields that are consisted of two counter-rotating pulses. This is maybe a reason why the spiral can exist even if the single field is present.

These results suggest that the spin effect of created particles plays an important role in momentum spiral on pair production. It can not only induce the momentum spiral in a single field, but also contribute to the formation of the spiral in the bichromatic combinational fields as well as the appearance of spin flip effect. Moreover, the momentum spiral due to the spin is sensitive to the field parameters. It is found that either the larger the time delay or the smaller the number of cycles, the more pronounced the momentum spiral, which provides a flexible way to control the spiral features. While we have only considered the simple case of CP fields, we speculate that the more delicate spin effects exist in different cases of fields and further features associated to the momentum patterns on pair production are expected.

\textit{Acknowledgements.}
We are grateful to ZL Li and C Kohlf\"urst for helpful discussions. This work was supported by the National Natural Science Foundation of China (NSFC) under Grant No.\ 12375240, No.\ 11935008 and No.\ 12104428. The computation was carried out at the HSCC of the Beijing Normal University. \


\begin{thebibliography}{100}

\bibitem{Sauter:1931zz}
F.~Sauter,
Uber das Verhalten eines Elektrons im homogenen elektrischen Feld nach der relativistischen Theorie Diracs,
Z.\ Phys.\  {\bf 69}, 742 (1931).

\bibitem{Heisenberg:1935qt}
W.~Heisenberg and H.~Euler,
Consequences of Dirac's theory of the positron,
Z.\ Phys.\  {\bf 98}, 714 (1936).

\bibitem{DiPiazza:2011tq}
A.~Di Piazza, C.~Muller, K.~Z.~Hatsagortsyan, and C.~H.~Keitel,
Extremely high-intensity laser interactions with fundamental quantum systems,
Rev. Mod. Phys. \textbf{84}, 1177 (2012).

\bibitem{Fedotov:2022}
A.~Fedotov, A.~Ilderton, F.~Karbstein, B.~King, D.~Seipt, H.~Taya, and G.~Torgrimsson,
Advances in QED with intense background fields,
Phys. Rept. \textbf{1010}, 1 (2023).

\bibitem{Xie:2017}
B. S. Xie, Z. L. Li, and S. Tang,
Electron-positron pair production in ultrastrong laser fields,
Matter Radiation. Extremes {\bf 2}, 225 (2017).

\bibitem{Dunne:2008kc}
G.~V.~Dunne,
New strong-field QED effects at ELI: nonperturbative vacuum pair production,
Eur. Phys. J. D \textbf{55}, 327 (2009).

\bibitem{Schwinger:1951nm}
J.~S.~Schwinger,
On gauge invariance and vacuum polarization,
Phys.\ Rev.\  {\bf 82}, 664 (1951).

\bibitem{Burke:1997ew}
D.~L.~Burke, R.~C.~Field, G.~Horton-Smith, J.~E.~Spencer, D.~Walz, S.~C.~Berridge, W.~M.~Bugg, K.~Shmakov, A.~W.~Weidemann, C.~Bula, K.~T.~McDonald, and E.~J.~Prebys,
Positron production in multi-photon light by light scattering,
Phys. Rev. Lett. \textbf{79}, 1626 (1997);
C.~Bamber, {\it et al.},
Studies of nonlinear QED in collisions of 46.6-GeV electrons with intense laser pulses,
Phys. Rev. D \textbf{60}, 092004 (1999).

\bibitem{Kohlfurst:2013ura}
C.~Kohlf\"urst, H.~Gies, and R.~Alkofer,
Effective mass signatures in multiphoton pair production,
Phys. Rev. Lett. \textbf{112}, 050402 (2014).

\bibitem{Hebenstreit:2011wk}
F.~Hebenstreit, R.~Alkofer, and H.~Gies,
Particle self-bunching in the Schwinger effect in spacetime-dependent electric fields,
Phys.\ Rev.\ Lett.\  {\bf 107}, 180403 (2011).

\bibitem{Kohlfurst:2017hbd}
C.~Kohlf\"urst and R.~Alkofer,
Ponderomotive effects in multiphoton pair production,
Phys. Rev. D \textbf{97}, 036026 (2018).

\bibitem{Li:2014En}
Z.~L.~Li, D.~Lu, B.~S.~Xie, L.~B.~Fu, J.~Liu and B.~F.~Shen
Enhanced pair production in strong fields by multiple-slit interference effectwith dynamically assisted Schwinger mechanism,
Phys. Rev. D \textbf{89}, 093011 (2014).

\bibitem{Li:2015Non}
Z.~L.~Li, D.~Lu, B.~S.~Xie, B.~F.~Shen, L.~B.~Fu, and J.~Liu,
Nonperturbative signatures in pair production for general elliptic polarization fields,
Europhys. Lett. \textbf{110}, 51001 (2015).

\bibitem{Li:2017qwd}
Z.~L.~Li, Y.~J.~Li, and B.~S.~Xie,
Momentum vortices on pairs production by two counter-rotating fields,
Phys. Rev. D \textbf{96}, 076010 (2017).

\bibitem{Hu:2023pmz}
L.~N.~Hu, O.~Amat, L.~Wang, A.~Sawut, H.~H.~Fan, and B.~S.~Xie,
Momentum spirals in multiphoton pair production revisited,
Phys. Rev. D \textbf{107}, 116010 (2023).

\bibitem{Li:2019hzi}
Z.~L.~Li, B.~S.~Xie, and Y.~J.~Li,
Vortices in multiphoton pair production by two-color rotating laser fields,
J. Phys. B \textbf{52}, 025601 (2019).

\bibitem{Ngoko:2015Ele}
J.~M.~Ngoko Djiokap, S.~X.~Hu, L.~B.~Madsen, N.~L.~Manakov, A.~V.~Meremianin, and Anthony F. Starace1,
Electron vortices in photoionization by circularly polarized attosecond pulses,
Phys. Rev. Lett. \textbf{115}, 113004 (2015).

\bibitem{Pengel:2017Elec}
D.~Pengel, S.~Kerbstadt, D.~Johannmeyer, L.~Englert, T.~Bayer, and M.~Wollenhaupt,
Electron vortices in femtosecond multiphoton ionization,
Phys. Rev. Lett. \textbf{118}, 053003 (2017).

\bibitem{Majczak:2022}
M.~M.~Majczak, F.~Cajiao V\'{e}lez, J.~Z.~Kami\'{n}ski, and K.~Krajewska,
Carrier-envelope-phase and helicity control of electron vortices and spirals in photodetachment,
Opt. Express \textbf{30}, 43330 (2022).

\bibitem{Macek2009}
J.~H.~Macek, J.~B.~Sternberg, S.~Y.~Ovchinnikov, T.~G~Lee,
and D.~R.~Schultz, Origin, evolution, and imaging of vortices in atomic processes, Phys. Rev. Lett. \textbf{102}, 143201 (2009).

\bibitem{K2016}
K.~J.~Yuan, S.~Chelkowski, and A.~D.~Bandrauk, Photoelectron momentum distributions of molecules in bichromatic circularly polarized attosecond uv laser fields,
Phys. Rev. A \textbf{93}, 053425 (2016).

\bibitem{L2020}
L.~Geng, F.~C.~V\'{e}lez, J.~Z.~Kami\'{n}ski, L.~Y.~Peng, and K.~Krajewska, Vortex structures in photodetachment by few-cycle circularly polarized pulses,
Phys. Rev. A \textbf{102}, 043117 (2020).

\bibitem{Harris1994}
M.~Harris, C.~A.~Hill, and J.~M.~Vaughan, Optical helices and spiral interference fringes, Opt. Commun.
\textbf{106}, 161 (1994).

\bibitem{Blatter1994}
G.~Blatter, M.~V.~Feigelman, V.~B.~Geshkenbein, A.~I.~Larkin, and V.~M.~Vinokur, Vortices in high-temperature superconductors, Rev. Mod. Phys. \textbf{66}, 1125 (1994).

\bibitem{Uby1995}
L.~Uby, M.~B.~Isichenko, and V.~V.~Yankov, Vortex filament dynamics in plasmas and superconductors, Phys. Rev. E \textbf{52}, 932 (1995).

\bibitem{Shukla2009}
P.~K.~Shukla and B.~Eliasson, {\it Colloquium}: Fundamentals of dust-plasma interactions, Rev. Mod. Phys. \textbf{81}, 25 (2009).

\bibitem{Madison2000}
K.~W.~Madison, F.~Chevy, W.~Wohlleben, and J.~Dalibard, Vortex formation in a stirred Bose-Einstein Condensate, Phys. Rev. Lett. \textbf{84}, 806 (2000).

\bibitem{Strobel:2014tha}
E.~Strobel and S.~S.~Xue,
Semiclassical pair production rate for rotating electric fields,
Phys. Rev. D \textbf{91}, 045016 (2015).

\bibitem{Blinne:2015zpa}
A.~Blinne and E.~Strobel,
Comparison of semiclassical and Wigner function methods in pair production in rotating fields,
Phys. Rev. D \textbf{93}, 025014 (2016).

\bibitem{Li:2019}
Z.~L.~Li, B.~S.~Xie, and Y.~J.~Li,
Boson pair production in arbitrarily polarized electric fields,
Phys. Rev. D \textbf{100}, 076018 (2019).

\bibitem{Kohlf2019S}
C.~Kohlf\"urst,
Spin states in multiphoton pair production for circularly polarized light,
Phys.\ Rev.\ D {\bf 99}, 096017 (2019).

\bibitem{Kohlfurst:2015zxi}
C.~Kohlf\"urst,
Electron-positron pair production in inhomogeneous electromagnetic fields,
Ph.D. thesis, arXiv:1512.06082 (2015).

\bibitem{Hebenstreit:2011pm}
F.~Hebenstreit,
Schwinger effect in inhomogeneous electric fields,
Ph.D. thesis, arXiv:1106.5965 (2011).

\bibitem{Bialynicki-Birula:1991jwl}
I.~Bialynicki-Birula, P.~Gornicki and J.~Rafelski,
Phase space structure of the Dirac vacuum,
Phys. Rev. D \textbf{44}, 1825 (1991).

\bibitem{Hebenstreit:2010vz}
F.~Hebenstreit, R.~Alkofer, and H.~Gies,
Schwinger pair production in space and time-dependent electric fields: Relating the Wigner formalism to quantum kinetic theory,
Phys. Rev. D \textbf{82}, 105026 (2010).

\bibitem{Ababekri:2019dkl}
M.~Ababekri, B.~S.~Xie, and J.~Zhang,
Effects of finite spatial extent on Schwinger pair production,
Phys.\ Rev.\ D {\bf 100}, 016003 (2019).

\bibitem{Kohlf2020Effect}
C.~Kohlf\"urst,
Effect of time-dependent inhomogeneous magnetic fields on the particle momentum spectrum in electron-positron pair production,
Phys.\ Rev.\ D {\bf 101}, 096003 (2020).

\bibitem{Blinne:2013via}
A.~Blinne and H.~Gies,
Pair production in rotating electric fields,
Phys. Rev. D \textbf{89}, 085001 (2014).

\bibitem{Li:2021vjf}
L.~J.~Li, M.~Mohamedsedik, and B.~S.~Xie,
Enhanced dynamically assisted pair production in spatial inhomogeneous electric fields with the frequency chirping,
Phys. Rev. D \textbf{104}, 036015 (2021).

\bibitem{Olugh:2018seh}
O.~Olugh, Z.~L.~Li, B.~S.~Xie, and R.~Alkofer,
Pair production in differently polarized electric fields with frequency chirps,
Phys. Rev. D \textbf{99}, 036003 (2019).

\bibitem{Li:2015cea}
Z.~L.~Li, D.~Lu, and B.~S.~Xie,
Effects of electric field polarizations on pair production,
Phys. Rev. D \textbf{92}, 085001 (2015).

\bibitem{Kohlfurst:2017git}
C.~Kohlf\"urst,
Phase-space analysis of the Schwinger effect in inhomogeneous electromagnetic fields,
Eur.\ Phys.\ J.\ Plus {\bf 133}, 191 (2018).

\bibitem{Olugh:2019nej}
O.~Olugh, Z.~L.~Li, and B.~S.~Xie,
Dynamically assisted pair production for various polarizations,
Phys. Lett. B \textbf{802}, 135259 (2020).

\bibitem{Dumlu1011}
C. K. Dumlu and G. V. Dunne, The stokes phenomenon and Schwinger vacuum pair production in time-dependent laser pulses, Phys. Rev. Lett. {\bf 104}, 250402 (2010).

\bibitem{Strobel2015}
J.~Oertel and R.~Sch\"{u}tzhold, WKB approach to pair creation in spacetime-dependent fields: The case of a spacetime-dependent mass, Phys. Rev. D {\bf 99}, 125014 (2019).

\bibitem{Kohlfurst:2022edl}
C.~Kohlf\"urst,
Pair Production in Circularly Polarized Waves, arXiv:2212.03180.

\end{thebibliography}
\end{document}